# Optical modes in linear arrays of dielectric spherical particles: A numerical investigation

Gail S. Blaustein,[b] Alexander L. Burin[*a]

[a] Department of Chemistry and [b]Department of Mathematics, Tulane University, New Orleans, LA, USA 70118-5636

## ABSTRACT

We have investigated bound modes in finite linear chains of dielectric particles of various lengths, interparticle spacing and particle materials. Through a unique application of the multisphere Mie scattering formalism, we have developed numerical methods to calculate eigen-optical modes for various arrays of particles. These numerical methods involve the use of the multisphere scattering formalism as the entries in $N \times N$ matrices where $N$ represents the number of particles in the chain. Eigenmodes of these matrices correspond to the eigen-optical modes of interest. We identified the eigenmodes with the highest quality factor by the application of a modified version of the Newton-Raphson algorithm. We found that convergence is strong using this algorithm for linear chains of up to several hundreds of particles. By comparing the dipolar approach with the more complex approach which utilizes a combination of both dipolar and quadrupolar approaches, we demonstrated that the dipolar approach has an accuracy of approximately *99%*. We found that the quality factor $Q$ of the mode increases with the cubed value of the number of particles in chain in agreement with the previously developed theory, the effects of disordering of particle sizes and inter-particle distances will be discussed.

## 1. INTRODUCTION

One-dimensional chains of scattering optical particles is attracting increasing interest because they can be used in various nano-applications where optical energy is received, transferred and converted on a subwavelength scale.[1-3] To our knowledge, the first practical nanowaveguide capable of transferring optical energy within a distance of *100* nm was developed by Atwater and coworkers[2,3] which was formed as a linear chain of spherical silver or gold particles (see Fig. 1). Excitation of the surface plasmon resonance in a particle at one end of the waveguide causes energy to be transferred to all particles in turn through the excitation of plasmon resonances in adjacent particles down the chain. The energy that is transferred to the other end of the waveguide stimulates any number of processes such as the photoexcitation of a molecule or a chemical reaction. The efficiency of energy transfer greatly decreases with the length of the waveguide due to cumulative energy loss.

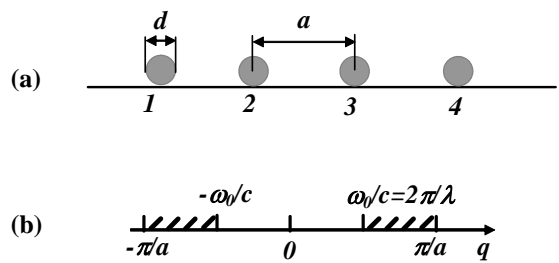

Figure 1. Linear chain of scattering particles (a) and possible values of the quasi-wavevector of propagating polariton modes with marked domain of guiding modes (b). All notations are described within the text.

[*] gblaustein@gmail.com; aburin@tulane.edu; phone (504) 862-3574; fax (504) 865-5596

Optical energy is also vulnerable to radiative losses associated with photon emission and absorption of light by the material. Losses due to photon emission can be almost completely suppressed if the energy used for particle excitation belongs to the energy domain of guiding modes. This energy domain exists where the photon emission process is forbidden by the momentum conservation law. In the infinite chain with the period *a* (Fig. 1), all optical excitations can be classified by their quasi-momentum projection to the chain axis *z*, which belongs to the domain *(-π/a, π/a)*.[4] If the resonant photon wavevector $k=\omega/c$ is less then the maximum excitation wavevector $\pi/a$, then two branches of guiding modes are formed

$$\left(-\frac{\pi}{a}, -\frac{\omega}{c}\right), \left(\frac{\omega}{c}, \frac{\pi}{a}\right), \quad (1)$$

where photon emission is forbidden by the momentum conservation law (see Fig. 1). The criterion $\omega/c < \pi/a$ is equivalent to the condition that the interparticle distance *a* should be less than the resonant wavelength, or

$$a < \frac{\pi c}{\omega} = \frac{\pi}{k} = \frac{\lambda}{2}. \quad (2)$$

This result is equivalent to the light cone constraint suggested in Ref. [5] and used later in Refs. [6, 7] to study long-living modes in chains of particles for various applications of interest.

It is straightforward to satisfy Eq. (2) for metal particles because their plasmon resonance frequencies correspond with the visible light spectrum as their wavelengths can be as long as few hundred nanometers with a particle diameter of as low as *30* nm, which is the minimum distance between particles. That is why metal particles can be used to build waveguides free of radiative losses to transfer optical energy on a subwavelength scale.[2,3]

However, all metals characteristically absorb optical energy due to the nature of their conducting electrons. That is to say they have continuous spectrum which can consume arbitrary amounts of photon energy thus strongly reducing the energy transmission in nano-waveguides.[2, 3] Therefore, it may be more convenient to use dielectric materials having very weak light absorption. Then, instead of relying on surface plasmon resonance, one can make use of Mie scattering resonances to form polariton modes where oscillations of material polarizations are coupled with electromagnetic waves. According to rudimentary estimates of Ref. [7] made within the framework of the simple dipolar oscillator model, guiding modes satisfying Eq. (2) can indeed be formed in chains of particles made of optical materials like $TiO_2$, ZnO or GaAs.[8] In contrast with metals, dielectric particles have negligible absorption of light so dielectric particle chains can transfer optical energy particularly efficiently.

Optical modes, or "polaritons," are collective excitations created from the superposition of material polarizations and photons. In this paper, we will study polaritons formed within long, finite chains of dielectric spherical particles possessing the longest possible lifetime (highest quality factor) and, for that reason, having the greatest relevance regarding energy manipulations using particle chains.

We will investigate the optical excitations within a one-dimensional chain of particles (Fig. 1) using the multi-sphere Mie scattering formalism[9, 10] which has a distinct advantage compared to the standard finite difference time domain (FDTD) approach[11] in that the Mie scattering formalism represents a scattering particle polarization generally characterized by an infinite number of parameters and a discrete set of multipole moments. Our solution for polariton modes is valid everywhere in space; therefore, our result is independent of boundary conditions that critically affect the FDTD method. As we show here, the analysis of the quality factor (not necessarily field amplitudes) of low energy optical modes for reasonably large refractive indices ($n_r > 2$) can be performed using a simple dipolar approach, which means that numerical calculations can be performed quickly.

This paper is organized as follows. In Section 2, we describe the multi-sphere Mie scattering formalism and its application to the investigation of eigensolutions of Maxwell's equations for the particle chain. In Section 3, the results for the quality factor of particle chains using the dipolar approach are presented. We also discuss the numerical results for quality factor dependence on the number of particles and the physical nature of obtained dependencies. In addition, we consider the effect of particle size disordering on the quality factor. In Section 4, we present our conclusions.

## 2. THE MULTISPHERE MIE SCATTERING FORMALISM IN THE STUDY OF PARTICLE CHAINS

The multisphere Mie scattering formalism has been developed to study the scattering of optical waves in aggregates of spherical dielectric particles. In our other work[12] for instance, we have used the multisphere Mie scattering formalism[9, 10, 13] to study quality factors of whispering gallery modes in circular arrays of spherical particles. This formalism uses the spherical vector function expansion of solutions to Maxwell's equations in the frequency domain taken at frequency $z$. The scattering wave partial amplitudes $a_{mn}^l$, $b_{mn}^l$ ($l = 1, 2, \ldots, N$) describing two transverse components with angular momentum $n$ and its projection $m$ to the chain axis all taken with respect to the center of the $l^{th}$ sphere are expressed through partial amplitudes $p_{mn}^l$ and $q_{mn}^l$ and by making use of matrices $A$ and $B$ defined by the vector translation coefficients

$$\frac{a_{mn}^l}{\bar{a}_n^l} + \sum_{j \neq l}^{(1,N)} \sum_{\nu=1}^{+\infty} \sum_{\mu=-\nu}^{\nu} \left( A_{mn\mu\nu}^{jl} a_{\mu\nu}^j + B_{mn\mu\nu}^{jl} b_{\mu\nu}^j \right) = p_{mn}^l,$$

$$\frac{b_{mn}^l}{\bar{b}_n^l} + \sum_{j \neq l}^{(1,N)} \sum_{\nu=1}^{+\infty} \sum_{\mu=-\nu}^{\nu} \left( B_{mn\mu\nu}^{jl} a_{\mu\nu}^j + A_{mn\mu\nu}^{jl} b_{\mu\nu}^j \right) = q_{mn}^l, \quad (3)$$

(see Refs. [9, 10] for details) where $\bar{a}_n^l$, $\bar{b}_n^l$ are Mie scattering coefficients for the $l^{th}$ sphere[9, 10], matrices $A$ and $B$ describe the multipole interactions, and cases $n=1$ or $\nu=1$ correspond to dipole moments, $n=2$ describes quadrupole moment, etc. It can easily be checked using definitions[9, 10] that the components $A_{m1\mu1}^{jl}$ describe the delayed dipole-dipole interaction of polarizations between two spheres. Eq. (3) is written in the frequency domain so all its coefficients depend on the frequency $z$ of the outgoing wave.

Since we are interested in the eigenmodes of a finite system, we set the left hand side of Eq. (3) describing the source term to zero, so we have

$$\frac{a_{mn}^l}{\bar{a}_n^l} + \sum_{j \neq l}^{(1,N)} \sum_{\nu=1}^{+\infty} \sum_{\mu=-\nu}^{\nu} \left( A_{mn\mu\nu}^{jl} a_{\mu\nu}^j + B_{mn\mu\nu}^{jl} b_{\mu\nu}^j \right) = 0,$$

$$\frac{b_{mn}^l}{\bar{b}_n^l} + \sum_{j \neq l}^{(1,N)} \sum_{\nu=1}^{+\infty} \sum_{\mu=-\nu}^{\nu} \left( B_{mn\mu\nu}^{jl} a_{\mu\nu}^j + A_{mn\mu\nu}^{jl} b_{\mu\nu}^j \right) = 0. \quad (4)$$

This approach is equivalent to setting the boundary condition equal to zero at infinity for outgoing waves. The homogeneous equation has a nontrivial solution only at a discrete set of frequencies thus making the determinant of the system equal zero. Generally, these solutions have a finite imaginary part because of the "dissipative" boundary conditions. To make the problem numerically solvable, we have to restrict the number of multipoles by some upper maximum index $n_{max}$. In this manuscript, we take the most straightforward approach $n_{max}=1$, which is equivalent to the dipolar approach.[14] A similar approach was taken in Refs. [8, 15-17] and here we report and discuss these results in greater detail.

As in Ref. [17], we also ignore the off-diagonal interaction $B$ between the $a_{m1}^l$ and $b_{m1}^l$ components and consider only the component $b_{m1}^l$ corresponding to the lowest Mie resonance

$$\frac{b_{m1}^l}{\bar{b}_1^l} + \sum_{j \neq l}^{(1,N)} \sum_{\mu=-1}^{1} A_{m1\mu1}^{jl} b_{\mu 1}^j = 0. \quad (5)$$

Recall that Mie scattering coefficients are defined by Ricatti-Bessel functions $\psi_1(x) = \sin(x)/x - \cos(x)$ and $\varsigma_1(x) = e^{ix}(-1 - i/x)$ as

$$\bar{b}_1 = \frac{n_r \psi_1(kd/2)\psi_1'(n_r kd/2) - \psi_1(n_r kd/2)\psi_1'(kd/2)}{n_r \varsigma_1(kd/2)\psi_1'(n_r kd/2) - \psi_1(n_r kd/2)\varsigma_1'(kd/2)}. \tag{6}$$

The nontrivial solutions of Eq. (5) exist for a discrete sequence of eigenfrequencies $z_a = \omega_a + i\gamma_a$. We are interested in systems of finite size where the emission of photons is inevitable so that each eigenmode must possess a finite decay rate $\gamma_a$.

Our consideration starts with the analysis of Eq. (5). This equation defines the spectral branch corresponding to Mie resonances associated with the zeros of the inverse Mie scattering amplitude $1/\bar{b}_1^l$. Since the second amplitude has a resonance at a lower frequency (cf. Ref. [18]) it is of greatest interest to us, which is the reason we will study the modes associated with the solution of Eq. (5).

Following from our estimates for circular arrays,[12] the dipolar approach is quite accurate in all cases of interest for the linear chain as well since estimates for multipole interactions up to $n_{max}=3$ have been considered. We found that if the refractive index exceeds *2*, then the corrections to the resonance frequency are less than *2%*. Since the circular array with a large number of particles is quite similar to the long chain of particles, we expect that our conclusions can be applied to the particle chain as well. Therefore, we shall restrict our consideration to the dipolar approach. Although our conclusion about the applicability of the dipolar approximation is valid for frequencies, decay rates and quality factors, it is not necessarily valid for electric and magnetic fields, which may show singular behavior particularly at points where particles touch each other.[19]

The study will proceed as follows. First we will use Eq. (4b) to find the lowest energy modes in a finite linear chain of identical, equally separated dielectric spheres composed of various optical materials of interest. Then we will investigate the disordering effect on the quality factor of the most bound polariton modes. We will also describe the method of our study which incorporates the modified Newton-Raphson method in determining solutions to the transcendental equations. This method is partially inherited from Refs. [8, 15], where lasing modes were investigated for interacting dipolar oscillators.

Because of the rotational symmetry of the chain with respect to the rotation around *z*-direction, the projection *m* of the excitation angular momentum to the *z*-axis is conserved and *m=-1*, *0*, or *1* can be selected to be constant. Formally, this results from the fact that interactions $A_{m1\mu 1}^{jl}$ for both centers of spheres *j* and *l* belonging to the *z*-axis are equal to zero if $m \neq \mu$. Therefore, Eq. (5) splits into three independent modes including two identical transverse *t* modes with *m=1*, or *-1* and polarization perpendicular to the chain and one longitudinal (*l*) mode characterized by the angular momentum projection *m=0* and polarized parallel to the chain. Then any mode in the chain of *N* ordered or disordered spheres with centers along the *z*-axis can be written as

$$\hat{M}(z)\mathbf{b} = 0, \tag{7}$$

where *M* is the matrix $N \times N$ extracted from Eq. (5). The diagonal elements of this matrix are inverse Mie scattering coefficients and its off-diagonal elements are defined by the matrix *A* of dipolar interactions between polarizations of different spheres. Vector *b* represents *N* expansion coefficients $b_{m1}^l$. The nontrivial solution of Eq. (7) exists when the matrix *M* has a zero eigenvalue. Thus, we need to find the value of *z* which makes one of the eigenvalues of matrix *M(z)* zero and possesses the minimum imaginary part. To use the Newton-Raphson algorithm, we need to choose an initial frequency. This value can be chosen as the Mie resonance frequency for the typical sphere. Then we can define the function *f(z)* as some eigenvalue of the matrix *M(z)* (cf. [8, 15]) that minimizes the absolute value of the imaginary part of *z* in the next iteration defined in the standard way

$$z_{n+1} = z_n - \frac{f(z)}{df(z)/dz}. \tag{8}$$

The stable point of this algorithm is realized when *f(z)=0*. Then Eq. (7) and consequently Eq. (5) have a nontrivial solution because the matrix *M* has one zero eigenvalue. Our method of iteration has been chosen to find the minimum eigenvalue in the standard manner, although we cannot prove rigorously that it always converges to the right eigenfrequency. Comparison of our method to the simplified situation where the Fourier transform method can be used to solve the problem[12] supports the validity of our results because we arrive at the same solution using both methods.

## 3. INVESTIGATION OF PARTICLE CHAINS

We begin our study with the analysis of the quality factor of ordered linear chains of identical, equally spaced particles within the framework of the dipolar approach (Eq. (5)), after which we consider the effect of disordering.

It is clear that the regime of most strongly bound polariton modes is realized when particles are closest to each other. This takes place when the distance between the centers of particles is equal to their diameter. Therefore, we can set $d=a$.

In numerical calculations, we have chosen the interparticle distance $a=2$. Since mode frequencies and decay rates are inversely proportional to the particle size, one can easily recalculate them for any size $a$ of interest. The results are presented using the size independent quality factor $Q$, which is defined as one-half of the ratio of the real and imaginary parts of a mode eigenfrequency

$$Q = \frac{\text{Re}(z)}{|\text{Im}(z)|} = \frac{\omega}{2\gamma}. \qquad (9)$$

Our calculations of the quality factor for different refractive indices and different modes are presented in Figs. 2, 3 and 4. Since there is no difference between two transverse modes because of the chain symmetry, data are shown for the transverse mode $m=1$ only. In all materials, quality factors of transverse modes exceed those for longitudinal modes at least for the large number of particles.

Because GaAs has the highest refractive index of the materials we studied, it is not surprising that modes with the highest quality factor are found in linear chains of GaAs particles. Accordingly, its Mie resonance is at the lowest frequency so the guiding criterion Eq. (2) is most easily realized there. One should note that for only *10* particles the quality factor of the transverse mode *t* already exceeds *5000*, while the quality factor is over $5\times10^6$ for *N>70*. It is clear that optical energy can be guided very efficiently through long chains of GaAs particles.

Table 1. Quality factors of dielectric particle chains for different materials. The results are given for the longitudinal mode *l* and one of two transverse modes *t*, because both transverse modes behave identically.

| Material | GaAs | | TiO$_2$ (rutile) | | ZnO | |
|---|---|---|---|---|---|---|
| $n_r$ | 3.5 | | 2.7 | | 1.9 | |
| Mode | t | l | t | l | t | l |
| Q for N=10 | 5630 | 3105 | 726 | 792 | 35 | 21 |
| Q for N=50 | 899690 | 333381 | 9886 | 8943 | 341 | 85 |
| k (a=d=2) | 0.843 | 0.945 | 1.068 | 1.204 | 1.406 | 1.583 |
| $\sigma$ for N=20 | 0.009 | n/a | 0.025 | n/a | 0.05 | n/a |
| $\sigma$ for N=50 | 0.0015 | n/a | 0.009 | n/a | n/a | n/a |

The dependence of the quality factor on the number of particles for GaAs is shown in Fig. 2. It is clear that both longitudinal and transverse modes are bound for GaAs because of its high refractive index $n_r=3.5$. The resonant photon wavevector for both modes is less that the maximum polariton wavevector $q=\pi/a=\pi/2\approx1.57$, so the guiding criteria Eqs. (1), (2) are satisfied for both modes. In Figs. 3, 4 the quality factor dependence on the the number of particles is shown for two other materials of interest including TiO$_2$ (rutile phase, $n_r=2.7$) and ZnO ($n_r=1.9$) respectively. The most bound modes in TiO$_2$ behave quite similarly to GaAs. They are both bound because their resonant wavelengths satisfy very well the guiding criterion Eq. (2). Accordingly, the quality factor approaches infinity in the limit $N\rightarrow\infty$. The quantitative characteristics of the particle array made of TiO$_2$ are given in Fig. 4 and included into Table 1 similarly to those for GaAs. Still only ten particles are needed to attain the quality factor exceeding one hundred.

The quality factor behavior is quite different in ZnO possessing the lowest refractive index. The resonant wavelength for the longitudinal mode there $\lambda=2\pi/k=3.97$ does not satisfy the guiding criterion Eq. (2) at the minimum interparticle distance $d=2$. Therefore all longitudinal modes are expected to be unbound within the framework of the dipolar approach. Note that the wavevector exceeds the threshold by less than *2%* in its absolute value. The very weak increase of the quality factor of longitudinal mode with the number of particles seen in Fig. 4 can be therefore understood as the consequence of its small difference from the threshold value. Since corrections to the transversal mode resonant wavevector k were estimated to be around 2%[12] it is not clear whether our preliminary conclusion about longitudinal modes in ZnO would remain valid after more accurate study.

The data in Figs. 2, 3, 4 are presented in the log-log format because the dependence of the quality factor on the numer of particles can be described by the power law. This contrasts with the circular array behavior, where the quality factor increases with the number of particles exponentially.[9, 7, 17, 12] On our opinion the losses are stronger in finite chains because they have sharp ends where the emission of light takes place.

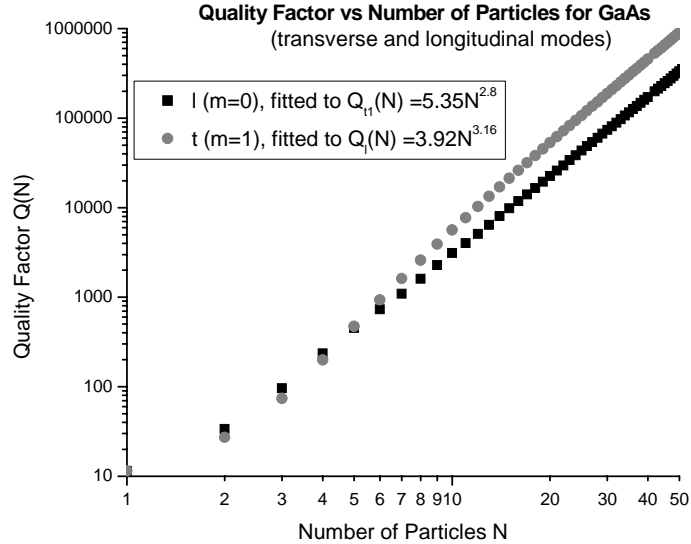

Figure 2. Quality factor dependence on the number of particles for GaAs ordered linear chains.

The quality factor dependence on the number of particles is close to the law

$$Q(N) \sim N^3 \qquad (10)$$

found previously in Ref. [8] in all materials except for ZnO, where the mode frequency for the transverse mode is very close to its threshold value Eq. (2) so the universal regime Eq. (2) is not probably reached there yet. The physical nature of this dependence has been interpreted based on the analysis of the lifetime of the mode in the finite system near the band edge, following the previous work by Dowling and coworkers.[20, 21] The main difference between this work and Ref. [17] is that here we are using the explicit Mie resonance following Ref. [16], while in Ref. [17] dielectric spheres were represented by harmonic oscillators. Therefore the situation is quite similar, but since the justification of $Q \sim N^3$ law was made very schematically in Ref. [17] we give more details here.

The most bound mode for all systems of interest is realized at the maximum polariton quasi-wavevector $q=\pi/a$. This was clearly demonstrated in Ref. [12] for circular arrays, where the modes are defined by their conserving quasi-angular momenta and the value of quasi-angular momentum for the most bound mode clearly corresponds to the maximum wavevector. There is no translational or rotational invariance in the finite chain, which can permit us to find exact momentum of the mode, but the most bound mode frequencies $\omega$ (wavevectors $k=\omega/c$) are nearly identical in circular arrays and linear chains (compare Tables in this work and in Ref. [12]). The point $q=\pi/a$ is the minimum or the maximum in the polariton energy band (cf. Eq. (1)) defined by the dispersion function $\omega(q)$ because of the relationship $\omega(\pi/a-x)= \omega(\pi/a+x)$ (cf. Refs. [17, 22]). Therefore the point $k=\pi/a$ corresponds to the top or the bottom of the energy band where the group velocity of the mode has the minimum leading to the maximum in its lifetime in the finite sample of the length $L$.

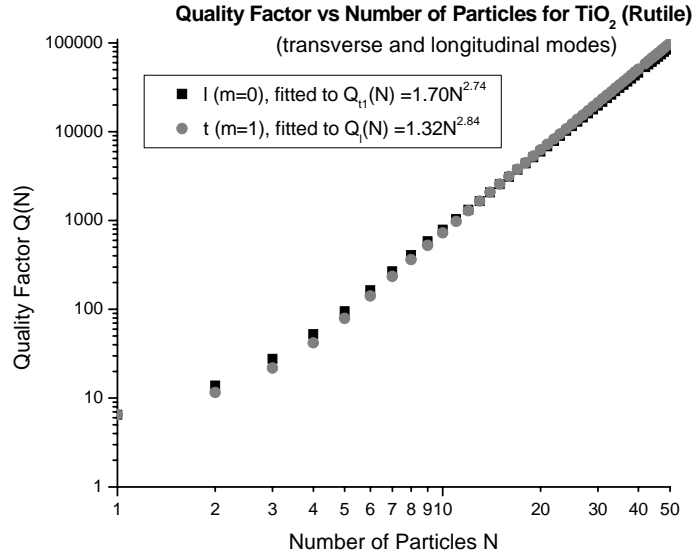

Figure 3. Quality factor dependence on the number of particles for TiO$_2$ ordered linear chains.

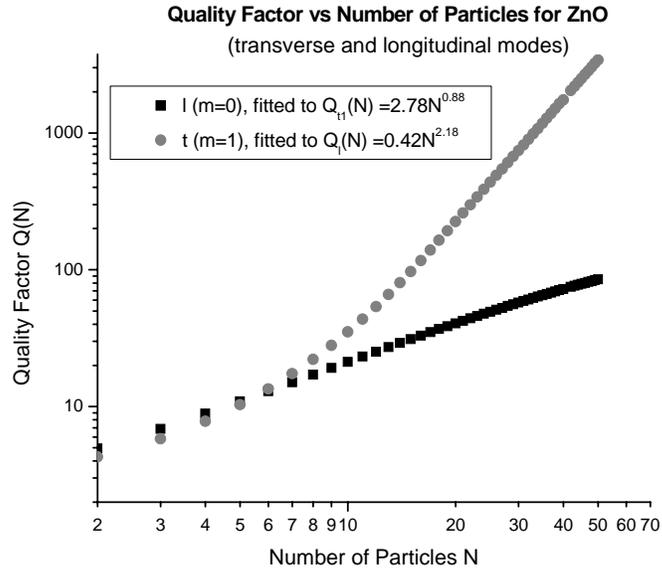

Figure 4. Quality factor dependence on the number of particles for ZnO ordered linear chains.

Following Ref. [17] we suggest the toy model to estimate the dependence of that lifetime (and the mode quality factor) on the system size $L$. In this model we consider the quantum mechanical quasistates of the particle with the unit mass within the material placed in the domain $(-L<x<L)$ and introducing there the potential energy $U_0>0$. The quantum states with the energy $E$ close to the potential minimum inside the material $E-U_0<<E$ are used to model the polariton modes near the band edge. Calculations performed in the Appendix shows that the quality factor of the mode with the lowest energy increase with the size of the system as (see Eq. (16) in Appendix)

$$Q \sim L^3 \tag{11}$$

which is equivalent to Eq. (10).

Finally we proceed to the discussion of disordering. In this paper we report the simple estimate of the maximum acceptable disordering necessary to reduce quality factor by the factor of *10*. We introduced disorder in particle sizes generating the random fluctuation in each particle diameter distributed normally with the certain dispersion $\sigma$. For the sake of simplicity particles were placed along the chain in touch with each other. A characteristic minimum value of $\sigma$ leading to the reduction of the quality factor by the factor of *M*, such as (*5<M<20*), was found by probing random values of *M* until the first success. This crude method must be quite sufficient for qualitative characterization of disordering effect. The dependence of the parameter $\sigma$ on the number of particles is shown in Fig. 7 for transverse modes in different materials.

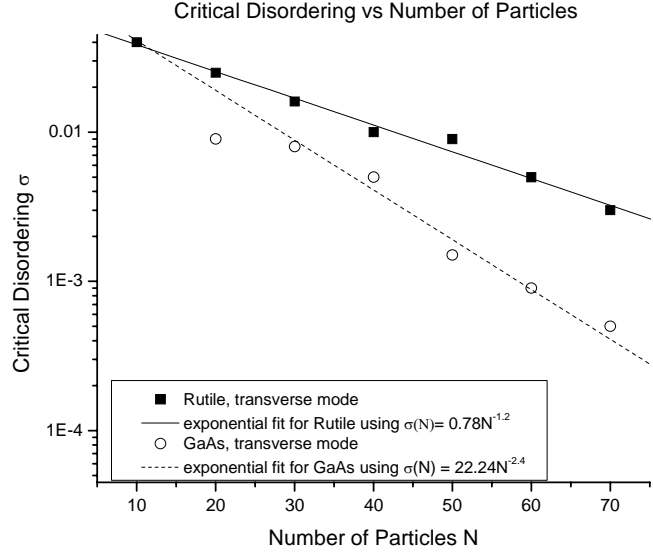

Figure 5. Acceptable disordering for GaAs and TiO$_2$, calculated assuming that it reduces the quality factor by around factor of ten compared to the regular chain of particles.

According to our crude modeling presented in Fig. 7 the "critical" dispersion $\sigma$ decreases with the number of particles *N* according to the power law $1/N^b$. This is not surprising because the quality factor increases according to the power law with *N* and the sensitivity of the system to disordering should increase in a correlated manner with increasing the quality factor. It is certainly not surprising that the critical disordering is smaller for GaAs, which has larger quality factor. The naïve expectation of the inverse proportionality of the critical disordering and the quality factor is definitely not confirmed by our study, so $\sigma$ decreases with the increase of *N* slower than *1/Q*. We cannot suggest any analytical relationship between the rates of changes for $\sigma$ and *Q*, because we do not have sufficient amount of data at this time.

## 4. CONCLUSION

In this paper we studied optical modes formed in one-dimensional finite chains of dielectric particles. Our main interest was whether such chain can possess the bound modes with the very high quality factors and what is the absolute value of the quality factor for particles made of different materials. We examined this question for three materials used in various optical applications including GaAs (refractive index $n_r=3.5$), TiO$_2$ ($n_r=2.7$, rutile phase) and ZnO ($n_r=1.9$). According to theoretical analysis it is possible to create bound modes within the chain having decay rates vanishing with the increase of the number of particles *N* and we tried to find out whether this can be done using standard optical materials.

To study optical modes we employed the multisphere Mie-scattering formalism used to find eigensolutions of Maxwell equations. We studied quasistates possessing the smallest decay rate or, equivalently, the highest quality factor.

Using the dipolar approximation we demonstrated that the quality factor of most bound polariton modes indeed increases with the number of particles within the chain for all three materials of interest approximately as $N^3$. It turns out that the most bound mode is the transverse mode with the polarization perpendicular to the circle plane and characterized by the quasi-momentum projection on the chain axis $q=\pi/a$ where $a$ is the period of the chain. The theoretical justification for the $N^3$ law has been suggested using the toy model of the modes near the band-edge of one-dimensional finite sample. The universal behavior $Q\sim N^3$ takes place in GaAs and $TiO_2$ having relatively high refractive indices, while ZnO having the refractive index $n_r=1.9$ is near the threshold for the formation of bound modes and therefore the quality factor behavior there is quite non-universal.

Disordering in particle sizes and placements remarkably reduces the quality factor of the system (See Table 1). The minimum acceptable disordering necessary to reduce the quality factor decreases with the number of particles as $1/N^b$ with the non-universal exponent b, which is however less than the exponent $3$ for the quality increase rate.

This work is supported by the U.S. Air Force Office of Scientific Research (Grant No. FA9550-06-1-0110). Authors acknowledge Olya Samoylova for writing the code to study disordering effect, performing numerical calculations, preparing graphs and deriving and writing the Appendix section. We are also grateful to Arthur Yaghjian, Svetlana Boriskina, Il'ya Polishchuk, Hui Cao and Alexei Yamilov for useful discussions and suggestions.

## 5. APPENDIX

We model the quality factor dependence on the number of particles using the mode having the longest life-time, which is the mode nearest to the band edge. The simplest model for such mode can be expressed using the one-dimensional Schrödinger equation with the potential

$$U(x) = \begin{cases} U_0, & -L < x < L, \\ 0, & \text{otherwise} \end{cases}$$

The quasistate with the lowest energy $E > U_0$ can be used to estimate the quality factor as

$$Q = -\frac{\operatorname{Re} E}{2 \operatorname{Im} E}.$$

We let the mass to be equal $1$ because it does not influence the quality factor dependence on the size $L$. Then one can represent the solution for the quasistate at $x > 0$ in the form

$$\Psi_0 = C_0 \cos(\sqrt{(E-U)}\, x), \quad x < L$$
$$\Psi_1 = C_1 e^{i\sqrt{E}\, x}, \quad x > L.$$

The boundary condition is selected as the outgoing wave as it is always supposed for quasistates.[23] Due to the symmetry of the model, it is sufficient to satisfy the boundary conditions at $x = L$, where the boundary conditions can be expressed as the continuity requirements for the wavefunction and its derivative

$$\Psi_0(L) = \Psi_1(L),$$
$$\Psi_0'(L) = \Psi_1'(L).$$

Making the appropriate substitutions yields the following system of equations

$$C_0 \cos(\sqrt{(E-U)}\, L) = C_1 e^{i\sqrt{E}\, L} \tag{12}$$

$$-\sqrt{(E-U)}\, C_0 \sin(\sqrt{(E-U)}L) = i\sqrt{E}\, C_1 e^{i\sqrt{E}L} \tag{13}$$

where Eq. (12) is for the wave function and Eq. (13) is for its derivatives. Dividing Eq.(12) by Eq.(13) gives us

$$-\frac{1}{\sqrt{(E-U)}\, \tan(\sqrt{E-U}\, l)} = -i E^{-\frac{1}{2}} \tag{14}$$

In our case, $L \to \infty$, $E \to U$, so in order for Eq. (14) to be valid, we must have

$$\sqrt{E-U} = a, \quad aL = \frac{\pi}{2} - \delta, \; \delta \ll L \tag{15}$$

Using Eq. (15) and $\tan \delta \approx \delta \approx i\frac{\pi}{2} U^{-\frac{1}{2}}$ we obtain

$$\frac{\pi}{2} - aL \approx i\frac{\pi}{2L}U^{-\frac{1}{2}}, \ a = \sqrt{E-U}, \ a \approx \frac{\pi}{2L}\left(1 - \frac{i}{L\sqrt{U}}\right)$$

$$a^2 = E - U \approx \frac{\pi^2}{4L^2}\left(1 - \frac{2i}{L\sqrt{U}}\right)$$

$$E \approx U + \frac{\pi^2}{4L^2} - i\frac{\pi^2}{2L^3\sqrt{U}}$$

Hence,

$$\mathrm{Im}\,E \approx -\frac{\pi^2}{2L^3\sqrt{U}}, \ \mathrm{Re}\,E = U + \frac{\pi^2}{4L^2} \approx U$$

and the quality factor can be expressed as[15]

$$Q = -\frac{\mathrm{Re}\,E}{2\,\mathrm{Im}\,E} = \frac{L^3}{\pi^2}U^{\frac{3}{2}}, \tag{16}$$

which correlates with our numerical findings.